\documentclass[aps,prl,twocolumn,superscriptaddress,footinbib]{revtex4}
\usepackage {amssymb}
\usepackage {amsmath}
\usepackage {graphicx}
\usepackage[dvipdfm]{hyperref}
\newcommand{\povo}{Department of Physics, University of Trento and INFN, Gruppo Collegato di
Trento, I-38050, Povo, Trento, Italy}
\newcommand{\fref}[1]{Fig.\ref{#1}}
\newcommand{\eref}[1]{Eq.(\ref{#1})}
\newcommand{\tref}[1]{Tab.(\ref{#1})}

\begin{document}

\title{Upper limits to surface force disturbances on LISA
proof-masses\\ and the possibility of observing galactic binaries}

\author{L. Carbone}
\affiliation{\povo}
\author{A. Cavalleri}
\affiliation{Istituto di Fotonica e Nanotecnologie CNR-ITC and INFN
Gruppo Collegato di Trento, I-38050 Povo, Trento, Italy}
\author{G. Ciani}
\affiliation{\povo}
\author{R. Dolesi}
\affiliation{\povo}
\author{M. Hueller}
\affiliation{\povo}
\author{D. Tombolato}
\affiliation{\povo}
\author{S. Vitale}
\affiliation{\povo}
\author{W. J. Weber}
\affiliation{\povo}

\begin{abstract}
We report on the measurement of parasitic surface force noise on a
hollow replica of a LISA (Laser Interferometer Space Antenna for the
observation of gravitational waves) proof-mass (PM) surrounded by a
faithful representation of its in-flight surroundings, namely the
capacitive sensor used to detect PM motion. Parasitic forces are
detected through the corresponding torque exerted on the PM and
measured with a torsion pendulum in the frequency range
$0.1\div30$~mHz. The sensor electrodes, electrode housing and
associated readout electronics have the same nominal design as for
the flight hardware, including 4~mm gaps around the PM along the
sensitive laser interferometry axis. We show that the measured upper
limit for surface forces would allow detection of a number of
galactic binaries signals with signal to noise ratio up to $\approx
40$ for 1 year integration. We also discuss how the flight test
under development, LISA Pathfinder, will substantially improve this
limit, approaching the performance required for LISA.
\end{abstract}

\maketitle

{\bf Introduction.} LISA is a space-borne gravitational wave
detector under development by the American (NASA) and European (ESA)
space agencies \cite{Vitale:2002}. The detector measures, by means
of a laser interferometer, the gravitational wave-induced change in
the distance between free-falling proof-masses (PMs) separated by
$\approx5 \times 10^9$m. Three PM pairs define three lines that form
a nearly equilateral triangle. The PM pairs are assumed to be in
pure geodesic motion, free of spurious \textit{relative}
acceleration noise along the measurement axis (referred to as $x$
from now on) to within $\sqrt{2}\times \{3\times 10^{-15} (
\textrm{m}/ \textrm{s}^{2}\sqrt{\textrm{Hz}} ) (1+(f/3 \textrm{mHz}
)^4)^{1/2}\}$, where $f$ is the frequency.

The PMs in LISA are cubes made of a Au-coated Au-Pt alloy and as
such are solid conducting bodies with no mechanical contact to the
surroundings. The relevant disturbing forces can be divided into
forces into those that act on the surface, which are most
significant, and those acting on the bulk, namely low frequency
magnetic and gravitational effects. Both the magnetic fields
generated inside the spacecraft (SC) and the PM magnetic properties
can be measured on ground with sufficient accuracy, and thus the
resulting magnetic force noise expected in flight can be well
characterized. The interplanetary magnetic field is known to be
stable enough to be a minor entry in the LISA error budget
\cite{Stebbins:2004}. As there are no continuously moving masses in
the current LISA design, the most worrisome gravitational noise
source is the SC thermo-mechanical distortion. Were this distortion
known in detail, the gravitational noise on the PM would then be
well modeled based on the measured mass properties. Modeling of
thermo-mechanical distortion is part of the LISA Pathfinder mission
\cite{Anza:2005}, the LISA precursor under development by ESA, as is
an in-flight measurement.

Surface forces include, among other disturbances, low frequency
electric field fluctuations, back-action from the electrostatic
position readout, thermal radiation effects, and residual pressure
gradients. These forces are the most difficult to model, as they may
be connected, for instance, to electrochemical or radiative
properties of the PM and electrode surfaces.

In LISA the PM is surrounded by a system of electrodes used to
detect the PM motion within the SC. 6 pairs of electrodes are used
to sense the proof mass motion in all degrees of freedom, using
100~kHz capacitance bridge readouts. In addition a set of biasing
electrodes on opposing faces along one or both directions normal to
$x$ provides the PM ac-voltage polarization required by the bridge
to work. The position information is fed back to control loops that
suppress relative SC-PM motion. We call this electrode system,
together with its support housing, the necessary readout coaxial
cables, and other accessories, the Gravity Reference Sensor (GRS).
The GRS creates a nearly closed cavity around the PM, and, with its
Au-coated electrodes and housing, serves as an electrostatic shield.
Thus the GRS shields the PM from external sources of surface force
and is, in practice, the dominant source of surface force
disturbances.

We proposed \cite{Dolesi:2000} a torsion pendulum to investigate
surface forces on LISA-like PMs. A first measurement campaign
\cite{Carbone:2003} employed a prototype sensor with a hollow, 40~mm
Au-coated Ti PM, suspended in-axis with a W wire, with 2~mm gaps to
the surrounding sensor electrode surfaces. The hollow PM allows use
of a thinner torsion fiber, and thus higher force sensitivity, than
with the full 2~kg LISA PM, without sacrificing representativity for
surface forces \cite{Dolesi:2000,Carbone:2003}. The electrodes were
Au-coated Mo plates, with biasing electrodes only along one axis
(the $z$-axis). A simple home-made sensor readout electronics was
used for the measurements.

In connection with the implementation of the LISA Pathfinder mission
\cite{Anza:2005}, the design of the GRS has evolved
\cite{Dolesi:2003}. The PM is now a 46~mm cube, separated from the
electrodes by a gap of 4~mm along the $x$-direction, and 2.9~mm and
3.5 mm on the other axes. The electrodes are now Au-coated ceramic
pieces mounted inside a Mo structure. In addition the biasing
electrodes are present on all faces except the ones along $x$.

In this article we describe the results of a measurement campaign
performed with the torsion pendulum on a prototype of this new
sensor design, connected to a readout electronics closely following
the final flight-hardware design. From the results we derive an
upper limit for surface forces on the PM for LISA. Finally, we
briefly compare this performance with that expected from LISA
Pathfinder for the entire force disturbance budget.

\begin{figure}[t]
\includegraphics[width=60mm]{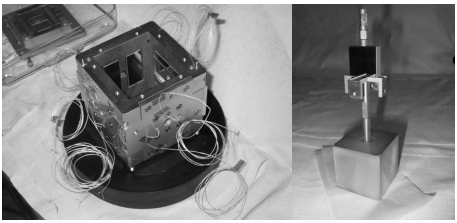}%{Foto1b.eps}
\caption{Left: the electrode housing of the GRS used for the
reported measurements. The AU-coated ceramics electrodes for the
x-direction motion readout are the pair of shiny vertical plates
visible inside the sensor. Also visible, on the adjacent y face, is
a y-sensing electrode, the upper horizontal plate, and one of the
electrodes used to polarize the PM for capacitive readout (with a
circular hole in the center). Right: the hollow PM used for the
torque measurements.} \label{fig1}
\end{figure}

The main features of the prototype are described above, with a
photograph shown in \fref{fig1}. Details of the readout scheme can
be found both in \cite{Carbone:2003,Dolesi:2003}. Also shown in
\fref{fig1} is the pendulum inertial element, featuring a 46~mm
hollow Al cube suspended by a W wire by means of an Al shaft. The
shaft is electrically isolated from the PM via a quartz ring hidden
underneath the top face of the PM. A mirror used for the independent
optical readout of the angular motion is attached to the shaft,
along with a structure used to balance the entire setup. The total
moment of inertia $I_o = \left( {4.31\pm 0.01} \right)\times 10^{ -
5}$~kg~m$^2$. The PM is suspended under high vacuum ($10^{-5}$ Pa).
The resulting torsional oscillator has a period approximately
$T_{o}=564$~s, the exact value depending on the applied electrical
field, and a quality factor $Q = 2900\pm 500$.

The GRS is mounted on a micro-manipulator that allows moving and
centering it around the PM, to which it has no mechanical contact.
This manipulator, together with an additional one at the upper
pendulum suspension point, allows the centering in all degrees of
freedom with $\mu $m precision. The pendulum motion is measured both
by the GRS and by an independent optical readout based on a
commercial autocollimator. The GRS has a sensitivity of order
2~nm/$\sqrt{\textrm{Hz}}$ in displacement and
200~nrad/$\sqrt{\textrm{Hz}}$ in rotation in rotation down to
frequencies of 1~mHz, with no reliable measurement below this
frequency. It provides the PM motion readout in all 6 degrees of
freedom. The autocollimator has an intrinsic resolution of
20~nrad/$\sqrt{\textrm{Hz}}$ within the same frequency range and
provides the readout for both the torsion angle and pendulum tilt
around one horizontal axis. The GRS is equipped with heaters and
thermometers, while optical fibers can shine UV light on PM or
electrode housing to discharge the PM. A 3-axis magnetometer
monitors the low frequency magnetic field fluctuations with a
resolution of 1~pT/$\sqrt{\textrm{Hz}}$ above 1~mHz and a $1/f$
upturn below this frequency. More details on the pendulum can be
found in \cite{Dolesi:2000,Carbone:2003}.

The angular motion of the pendulum $\phi (t)$ is converted into an
instantaneous external applied torque $N(t)$ as
\begin{equation}\label{eq1}
N\left( t \right) = I_o \left\{ {\ddot {\phi }\left( t \right) +
\frac{2\pi }{T_o Q}\dot {\phi }\left( t \right) + \left( {\frac{2\pi
}{T_o }} \right)^2\phi \left( t \right)} \right\}
\end{equation}
%\[
%N\left( t \right) = I_o \left\{ {\ddot {\phi }\left( t \right) +
%\left( {{2\pi } \mathord{\left/ {\vphantom {{2\pi } {T_o Q}}}
%\right. \kern-\nulldelimiterspace} {T_o Q}} \right)\dot {\phi
%}\left( t \right) + \left( {{2\pi } \mathord{\left/ {\vphantom
%{{2\pi } {T_o }}} \right. \kern-\nulldelimiterspace} {T_o }}
%\right)^2\phi \left( t \right)} \right\} \quad 1
%\]
\noindent In practice, the derivatives are estimated from a sliding
second order fit to 5 adjoining data. \eref{eq1} approximates the
damping as viscous, while in reality it is dominated by structural
dissipation within the fiber. However, the difference between these
two models has no detectable effect on data processing.

In order to get an estimate of the expected readout background
noise, we formed the average, $\bar{N}(t) =\left[ N_{GRS}(t)+
N_{ac}(t) \right]/ 2 $, and semi-difference, $\delta N(t)=\left[
N_{GRS}(t) - N_{ac}(t)\right]/2$ of the torques, $N_{GRS}(t)$ and
$N_{ac} (t)$, calculated, respectively, from the GRS and
autocollimator angular readouts $\phi _{GRS} \left( t \right)$ and
$\phi _{ac} \left( t \right)$. For ideally uncorrelated readout
noise, the power spectral densities (PSD) of these signals should
be:
\begin{eqnarray}
\label{eq2} S_{\bar{N}} \left( f \right) &=& S_N \left( f \right) +
\frac{S_{n,\phi_{GRS}}\left( f \right) + S_{n,\phi_{ac}}\left( f
\right)}{4\left| H\left( f \right)\right|^2}\\
\label{eq3} S_{\delta {N}} \left( f \right) &=&
\frac{S_{n,\phi_{GRS}}\left( f \right) + S_{n,\phi_{ac}}\left( f
\right)}{4\left| H\left( f \right)\right|^2}
\end{eqnarray}
\noindent where $S_X \left( f \right)$ is the PSD of process $X$ at
frequency $f$, $S_{n,\phi _{GRS} } \left( f \right)$ and $S_{n,\phi
_{ac} } \left( f \right)$ are the PSDs of the uncorrelated angular
readout noise of, respectively, the GRS and autocollimator, and
$H\left(f\right)=\{ I_o \left(2\pi\right)^2 [ \left(1/T_o\right)^2 -
f^2 + i \left(f/T_o Q\right) ]\}^{-1}$ is the pendulum
torque-to-angle transfer function. Thus within this model one can
estimate the torque PSD by:
\begin{equation}\label{eq4}
S_{N}\left(f\right)={S_{\bar{N}}\left(f\right) - S_{\delta
N}\left(f\right)}={\rm Re}\left\{ S_{N_{ac},N_{GRS}}\left(f\right)
\right\}
\end{equation}

\eref{eq4} indicates that in this model the difference $S_{\bar {N}}
\left( f \right) - S_{\delta N} \left( f \right)$, attributed to
true torque noise measured above the readout limit, is equivalent to
the correlated noise between the two readout signals, and can be
calculated as the cross-power spectral density (CSD) ${\rm
Re}\left\{ {S_{N_{ac} ,N_{GRS} } \left( f \right)} \right\}$ between
$N_{GRS} \left( t \right)$ and $N_{ac} \left( t \right)$. Note that,
in the absence of a definite excess torque, the result of this
estimate, if %whether
calculated as a PSD difference or CSD, can give
negative values due to statistical fluctuations.

PSD and cross-spectral densities are estimated by standard Welch
periodogram, using a Blackman-Harris 3$^{rd}$ order window and by
averaging over 20 data segments, adjoining data segments overlapping
by 50{\%}. Data segments are individually de-trended by subtracting
the best fit straight line to the data. The accuracy of the method
has been tested with a variety of simulated data. Simulation shows
that in the presence of large drift or in general substantial noise
power below the measurement bandwidth, the two lowest frequency
points follow the frequency shape of the side lobes of the spectral
window. These points are then discarded during further data
processing.
\begin{figure}[t]
\includegraphics[width=70mm]{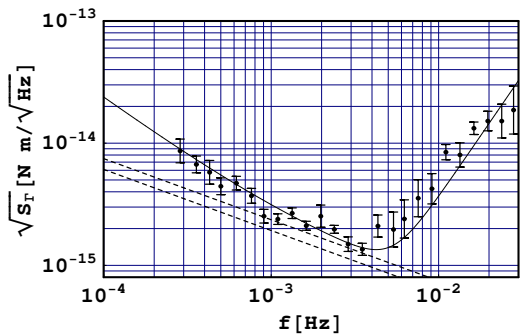}%{Fig1UpperLimit2.eps}
%\centerline{\includegraphics[width=12.57in,height=8.23in]{UpperLimit2.eps}}
\caption{Total torque noise. Data points: square root of torque
noise PSD estimated from autocollimator and capacitive sensor
cross-correlation after time domain subtraction of the effect of PM
motion along $y$. Values and error bars are estimated as the mean
and its standard deviation of PSD data falling within the
corresponding frequency bins. Bins have equal logarithmic width.
Dashed lines delimit the error band for the estimate of thermal
noise. The continuous smooth line is a linear least square fit to
data as described in the text.} \label{fig2}
\end{figure}
In \fref{fig2} we report an example of the outcome of this data
reduction procedure.

In this specific run, following an air re-entry into the vacuum
chamber, an extra coupling to motion of PM in the $y$-direction was
detected. $\bar{N}\left( t \right)$ data consequently showed a very
strong correlation with this motion as measured by the GRS. This
motion is a few orders of magnitudes larger than any SC motion
relevant for LISA, and thus a small coupling may convert into a
large torque that would not translate into a corresponding force on
a LISA PM. We thus subtracted the $y$-signal of the GRS multiplied
just by a constant from the $\bar {N}\left( t \right)$ data series.
The constant was chosen by minimizing the residual noise at low
frequency and was found to be $\partial \Gamma /\partial y =\left( -
9.8\,\pm 2.2\right) $~nN. This number, confirmed in other runs,
agrees with the results of coarse measurement obtained by moving the
GRS with the micromanipulator that gives $\partial \Gamma /\partial
y \approx7$~nN. Coincident with the appearance of this extra
coupling, additional angular stiffness was detected, measured by
rotating the GRS \cite{Carbone:2006}. This extra angular stiffness
amounts to $\partial^2 \Gamma /\partial \phi^2=(1.43 \pm 0.02)\times
10^{-10}$~Nm/rad. The data for ${\partial \Gamma }/{\partial y}$ and
for ${\partial ^2\Gamma } /{\partial \phi ^2}$ taken together,
suggest a localized charged feature on the $y$-face at $ \approx
1.5$~cm from the center, compatible with the observed slow decay of
the effect with time. We suspect that such a charge is due to one of
many observed defects in the electrode Au-coating that expose the
potentially-charged dielectric underneath. The overall effect of
subtraction is to decrease the noise by a factor 2 between 0.2~mHz
and $ \approx 2$~mHz, with negligible effect at higher frequencies.

The figure also shows the expected contribution from the thermal
torque noise associated with pendulum damping and providing the
ultimate torque noise floor. As damping at these low frequencies is
dominated by dissipation within the fiber, thermal noise is expected
to have a PSD
\begin{equation}\label{eq5}
S_{th} \left( f \right) = 4k_B T \frac{I_o \left(2\pi/{T_o
}\right)^2}{2\pi f Q}
\end{equation}
\noindent with $T$ the temperature and $k_{B}$ the Boltzmann
constant.

The figure shows that even after subtraction of the effect of motion
along $y$, residual noise coincides with the expected thermal
background, within the errors, only near 1-2~mHz. In order to
quantify the extra noise at other frequencies we performed a fit to
the excess with a simple polynomial law of frequency in the range
$0.2\div30$~mHz by a weighted least square procedure. Many different
power laws were tested with linear combinations of both positives
and negative exponents. The only choice giving fully significant
F-test value for all terms in the polynomial is the superposition of
a $1/f^{2}$ law with a term proportional to $f^{4}$. This is true
for all other experimental runs we have investigated (see below).
The results of the fitting are reported in \fref{fig2}. Coefficients
and uncertainties are reported in \tref{tab1}. The same table
reports the results for a run performed with the 40~mm PM, 2~mm gaps
prototype. Both runs were performed with same W torsion fiber.
\begin{table}[t]
\caption{Results of the estimate of excess noise
coefficients.}\label{tab1}
\begin{tabular}
{|p{52pt}|c |c |}%{|p{82pt}|p{83pt}|p{147pt}|p{129pt}|}
\hline \textbf{Prototype}& \textbf{f}$^{ - 2}$\textbf{coeff.}
[$\rm{N}^2\rm{m}^2\rm{Hz}$]&
\textbf{f}$^{4}$\textbf{coeff.} [${\rm{N}^2\rm{m}^2}{\rm{Hz}^{-5}}$] \\
%& & [$\rm{N}^2\rm{m}^2\rm{Hz}$] &
% [$\frac{\rm{N}^2\rm{m}^2}{\rm{Hz}^5}$]\\
\hline {\small 46~mm PM 4~mm gap}&  $\left( {3.7\pm 0.3} \right)
\times 10^{ - 36}$&
$\left( {3.8\pm 0.4} \right)\times 10^{ - 21}$ \\
\hline
 {\small 40~mm PM 2~mm gap} & $\left( {5.0\pm 0.4} \right)\times  10^{ -
36}$& $\left( {1.3\pm
0.2} \right) \times 10^{ - 21}$ \\
\hline
\end{tabular}
\label{tab1}
\end{table}

{\bf Discussion.} \fref{fig2} shows that excess noise only becomes
negligible around 2-3~mHz. Below this frequency the $1/f^{2}$ term
is a significant contribution to the noise that brings total noise
to a value which is 40{\%} above the background at 1~mHz and 2 times
this one at 0.2~mHz. We have not yet understood the origin of this
excess. There is no significant correlation with the magnetometer
readout. A weak, barely significant correlation between $\bar
{N}\left( t \right)$ and the thermometer monitoring the vacuum
column enclosing the torsion fiber was found. This temperature was
also correlated with the vertical motion $z(t)$ of the PM that
consequently showed a correlation with $\bar {N}\left( t \right)$. A
subtraction of this effect would lower the excess by only 10{\%} .

A few considerations seem to indicate that the excess is a property
of the pendulum and is not connected to the interaction between the
PM and GRS. First the noise level has not been significantly
affected by the change in the design of the GRS (\tref{tab1}), with
the difference between the results obtained with the two prototypes
being well within the variability, of a factor $ \approx 2$,
observed for each prototype among different runs taken at different
times during the many-month campaigns performed in each case. This
is true even though changing 2~mm to 4~mm gaps represents a major
change, significantly reducing, for instance, the coupling between
charge patches \cite{Speake:2003} and significantly increasing the
gas molecular conductance around the PM.

A candidate source for the measured torque excess is connected with
the continuous unwinding of the fiber due to creep in the fiber
material. In the runs presented, the fiber unwound with a rate of $
\approx  2\times 10^{ - 10}$~rad/s. The rate is reduced by
temperature annealing the fiber under load. Unfortunately we are
limited by practical reasons to an annealing temperature of only $
\approx 60^o$C. It is interesting to note that a simple Poisson
model where single angular slip events of amplitude $\delta \phi $
contribute to both unwinding and the residual noise, yields a step
size of $\delta \phi  \approx 40$~nrad and a rate of $ \approx  5
10^{ - 3}$~s$^{ - 1}$.

We searched for evidence of such glitches by running a
Wiener-Kolmogorov filter on the data matched to step-shaped torque
signals. The filter resolution for angular steps, resulted to be
$\sigma =33$~nrad, roughly the size expected from the reasoning
above. The filter did indeed detect several well identified large
amplitude ($3-5 \sigma $) events, satisfying goodness-of-the-fit
test criteria, and which accounted for $ \approx 20{\%}$ of the
excess noise at low frequency.

One can even entirely suppress the excess noise at low frequency by
subtracting the $ \approx $~200 highest events, with amplitude as
low as $\pm 2\sigma $. However, while these results can be taken as
an indication of angular glitches as the possible source for the
detected excess noise at low frequency, they cannot be considered as
an unambiguous proof and such events have not been subtracted from
the data of \fref{fig2}. It is worth remembering that any data
series can be approximated with arbitrary precision by a sequence of
variable amplitude steps and thus a subtraction procedure can
possibly, and erroneously, cancel the noise entirely.

The high frequency excess proportional to $f^{4}$ is also
statistically significant. The excess is due to a residual
correlation between the readouts at high frequency. The real part of
cross-coherence $r(f)={\rm Re} \left( {S_{\phi _{ac},\phi _{GRS}}
(f)}\right)/\sqrt{{S_{\phi _N } \left( f \right)S_{\phi _r } \left(
f \right)}}$, is still $0.10\pm 0.02$ at 10 mHz, dropping to
non-significant values only above 60 mHz.

It would be truly difficult to attribute this excess to any real
torque on the PM. For instance, when converted to a voltage on one
of the electrodes, this would amount to a PSD of 4 V/$\sqrt{ {\rm
Hz}}$ at 300~mHz assuming an unrealistically large dc voltage of 2 V
on the same electrode, all figures unlikely by orders of magnitude
\cite{Carbone:2006}.

The $f^{4}$ dependence converts into a white rotational noise of the
entire apparatus relative to the local frame of inertia by $ \approx
$30~nrad/$\sqrt{ {\rm Hz}}$. Thus a possible explanation may be such
a rotation due either to seismic noise or to local mechanical
distortion. However a residual electrical cross-talk between GRS and
autocollimator cannot be excluded.

Though the detected excess is likely not due to the GRS itself and
is not then relevant to LISA, we still include it when the
estimating the upper limit to the disturbances acting on the PM.
This upper limit also includes the fit uncertainty and by the
uncertainty in the evaluation of thermal noise, conservatively
estimated to be 20{\%} and coming from the error in Q-factor
measurement.

Overall this upper limit for the data of \fref{fig2}, defining
$f_{o}$= 1mHz is: {\small
\begin{eqnarray}
 3\cdot 10^{ - 30}\frac{{\rm N}^2{\rm m}^2}{Hz} \left\{ {1.2\left( {\frac{{f_o}}{f}}
\right)^2 + 0.3\left(\frac{{f_o}}{f} \right)}+  1.2\cdot 10^{ -
3}\left(
\frac{f}{{f_o}} \right)^4 \right. \nonumber \\
 \left. { + \sqrt {10^{ - 2}\left(
{\frac{{f_o}}{f}} \right)^4 +
 2.2\cdot 10^{ - 8}\left( {\frac{f}{{f_o}}} \right)^8}
} \right\}\nonumber
 \end{eqnarray}}
\noindent These torque data can be converted into an equivalent
acceleration noise with a few simple assumptions on the contributing
effects \cite{Carbone:2003}. For instance simulations
\cite{Carbone:2003} give that randomly distributed normal forces,
like those associated with parasitic electric fields, would exert
both torque and force noise, with a ratio of their amplitudes of
roughly $0.4 L$, with $L$ the PM size. This holds for disturbances
with spatial exponential correlation lengths up to $L/10$. The ratio
would decay to $L/4$ for a correlation slightly larger than $L$.
$L/4$ is also roughly the ratio obtained for an electrical
disturbance directly applied to the electrodes of the $x$-face, as
would be expected from electronic readout back-action. Finally
forces acting tangentially, like diffuse scattering of molecules,
show a ratio of $L/2$. We then convert the torque to an equivalent
acceleration noise that the corresponding forces would exert on the
1.96~kg LISA PM by using the lowest ratio $L/4$, i.e. 10.7~cm. The
reader should be aware though that obviously our pendulum geometry
would not detect disturbances applied uniformly and normal across
the $x$ face or tangentially to the $y$ and $z$ faces, or
disturbances acting normally onto the center of the $x$-face or in a
few other symmetric positions.

\begin{figure}[t]
%\centerline{\includegraphics[width=12.39in,height=8.05in]{UpperLimit3.eps}}
\includegraphics[width=70mm]{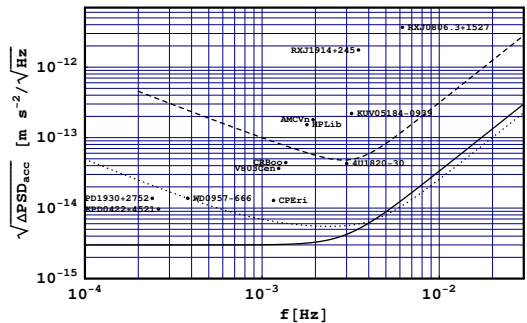}%{Fig2UpperLimit2b.eps}
\caption{\label{fig3}Dashed line: upper limit to pendulum torque
noise with GRS converted into an equivalent acceleration on one of
LISA PMs. Dots: acceleration level required to detect with LISA,
with signal to noise ratio of 1, the signal from the corresponding
galactic binary system over one year integration time. Dotted line:
projected upper limit for non-modeled overall force disturbances
that the LISA Pathfinder mission may achieve.}
\end{figure}

The results of the conversion of the data of \fref{fig2} into an
equivalent acceleration noise are reported in \fref{fig3}. Within
the same figure we also report for comparison the acceleration level
that would guarantee to LISA a measurement of the gravitational wave
signal from a few binary systems in our Galaxy over one year
integration time. The signal amplitude is taken from
\cite{Sterl:2002} and the LISA sensitivity is averaged for source
inclination and polarization.

It is important to stress again the limits of this observation.
Besides the lack of sensitivity to volume forces, all effects due to
coupling among different degrees of freedom that would take place
for a fully free-falling PM are absent here. Additionally the
environment on board LISA will be substantially different from the
laboratory one, with higher PM charging rate due to cosmic rays but
with lower level of temperature and magnetic field fluctuations. The
effects of these disturbances, however are well understood, have
been investigated separately by dedicated experiments
\cite{Carbone:2006}, and found in good agreement with the
expectations.

In \fref{fig3} we also report the projected result of a comparable
reduction process on the data from the LISA Pathfinder mission
\cite{Anza:2005}. The mission will measure, by means of a laser
interferometer, the relative acceleration of two LISA-like PMs in
free-fall within the same SC along the line joining their centers.

Mission requirements call for a demonstration of relative
acceleration of better than $3\times 10^{-15} ( \textrm{m}/
\textrm{s}^{2}\sqrt{\textrm{Hz}} ) (1+(f/3 \textrm{mHz} )^4)^{1/2} $
for frequencies between 1~mHz and 30~mHz. The relaxation of
requirements relative to LISA is dictated by the need of keeping the
test reasonably simple, but does not affect the GRS design. The
current estimate of the error budget \cite{Science:2005} places the
sensitivity somehow better than requirements at $ \approx 2\times
10^{ - 14}$ m/s$^2$$\sqrt{\rm Hz}$, with the largest contributions
being the cross-talk with motion of the PMs along degrees of freedom
other than the measurement ones, the magnetic field fluctuations,
and the residual coupling to the SC to gravitational and electrical
field gradients. This last coupling, which will be enhanced to allow
measuring the effect, can also be balanced between PMs in order to
suppress the influence of the SC coupling on their \textit{relative}
motion. PM motion along all degrees of freedom is measured with
sufficient precision to allow for a subtraction similar to what we
did with tilt in the case of the pendulum. LISA Pathfinder also
carries magnetometers so that magnetic field effect correction
should be possible. If these largest entries are suppressed the way
we have described here, LISA Pathfinder may be able to put an upper
limit on non-modeled disturbances to on a single PM acceleration of
$ \approx 5\times 10^{ - 15}$ m/s$^2$$\sqrt{\rm Hz}$ at 1~mHz. Above
this frequency the laser readout noise is expected to limit the
sensitivity, while below 1~mHz noise the degradation is expected to
be no worse than a $1/f^{2}$ dependence. This is the rationale
behind the curve shown in \fref{fig3}.

In conclusion current estimates of parasitic forces would indicate
that, even at the present level of experimental knowledge, LISA
should be able to detect bright gravitational wave signals. If LISA
Pathfinder will meet its expected performance, the risk of a major
unpredicted force disturbance will almost be entirely retired and
LISA would have basically its entire planned science objectives
guaranteed.

\end{document}